\documentclass[prd,nofootinbib,superscriptaddress,notitlepage,10pt]{revtex4-1}
\usepackage{epsfig,amsmath,amssymb,slashed}
\usepackage[utf8]{inputenc}
\usepackage{hyperref}
\usepackage{amsthm}
\usepackage{xcolor}
\usepackage{bm}
\usepackage[normalem]{ulem}

\theoremstyle{definition}

\newcommand{\di}{{\rm d}}

\newcommand{\e}{{\rm e}}

\newcommand{\be}{\begin{equation}}
\newcommand{\ee}{\end{equation}} 
\newcommand{\sqrts}{\sqrt{s_\mathrm{NN}}}

\begin{document}

\title{$\Lambda$ polarization in very high energy heavy ion collisions as a probe of the Quark-Gluon Plasma formation and properties}
\author{Andrea Palermo}
\affiliation{Center for Nuclear Theory, Department of Physics and Astronomy, Stony Brook University, Stony Brook, New York 11794-3800, USA}
\author{Eduardo Grossi}\affiliation{Universit\`a di 
 Firenze and INFN Sezione di Firenze, Via G. Sansone 1, 
	I-50019 Sesto Fiorentino (Florence), Italy}
\author{Iurii Karpenko}
	\affiliation{Faculty of Nuclear Sciences and Physical 
Engineering, Czech Technical University in Prague,\\  B\v rehov\'a 7, 11519 Prague 1, Czech Republic}
\author{Francesco Becattini}\affiliation{Universit\`a di 
 Firenze and INFN Sezione di Firenze, Via G. Sansone 1, 
	I-50019 Sesto Fiorentino (Florence), Italy}

\begin{abstract}
We have studied the spin polarization of $\Lambda$ hyperons in heavy ion collisions at center-of-mass 
energies $\sqrts = 200$ GeV and $\sqrts = 5.02$ TeV carried out at RHIC and LHC colliders. We have calculated 
the mean spin vector at local thermodynamic equilibrium, including all known first-order 
terms in the gradients of the thermo-hydrodynamic fields, assuming that the hadronization hypersurface 
has a uniform temperature. We have also included the feed-down contributions to the polarization of $\Lambda$ 
stemming from the decays of polarized $\Sigma^*$ and $\Sigma^0$ hyperons. The obtained results are in good agreement with the data. 
In general, the component of the spin vector along the global angular momentum, orthogonal to the reaction 
plane, shows strong sensitivity to the initial longitudinal flow velocity. Furthermore, the longitudinal 
component of the spin vector turns out to be very sensitive to the bulk viscosity of the plasma at the 
highest LHC energy. Therefore, the azimuthal dependence of spin polarization can effectively
constrain the initial hydrodynamic conditions and the transport coefficients of the Quark Gluon Plasma.
\end{abstract}

\maketitle

\section{Introduction}
After its measurement by STAR collaboration in 2017 \cite{STAR:2017ckg}, spin polarization has become 
an important probe in relativistic heavy ion collisions (for reviews, see 
\cite{Becattini:2020ngo,Becattini:2022zvf,Becattini:2024uha}).
From a theoretical standpoint, the most successful approach is the hydrodynamic-statistical model
where spin polarization is calculated at local thermodynamic equilibrium at hadronization when
the Quark Gluon Plasma (QGP) gives rise to hadronic particles which rapidly decouple and freely stream
to the detectors. 

In the local equilibrium model, the sources of the spin 
polarization vector of fermions are the gradients of the hydro-thermodynamic fields, that is, temperature, 
velocity, and chemical potential. Even if, for some time, spin polarization was mostly connected to 
vorticity (more precisely, thermal vorticity 
\cite{Becattini:2013fla,Fang:2016vpj,Florkowski:2018ahw,Weickgenannt:2020aaf}), it has become recently 
clear that also the symmetric gradient of the four-temperature (thermal shear tensor) and the gradient 
of the chemical potential \cite{Becattini:2021suc,Liu:2021uhn,Yi:2021ryh} are responsible for a large 
contribution to local polarization, even though the global polarization is essentially determined by 
the thermal vorticity \cite{Alzhrani:2022dpi}.

Particularly in very high energy collisions where the chemical potentials are negligible and the hadronization 
hypersurface $\Sigma$ can be approximated by an isothermal one, the mean spin polarization vector of 
Dirac fermions such as the $\Lambda$ hyperons with momentum $p$ is given by \cite{Becattini:2021iol} 
\footnote{It is important to point out that the formula \eqref{eq:polarization isothermal} somewhat differs
from others that have been used in other numerical studies 
\cite{Fu:2021pok,Wu:2022mkr,Jiang:2023fai,Jiang:2023vxp,Ribeiro:2023waz}, for a twofold reason. First, 
the equation \eqref{eq:polarization isothermal} is appropriate for an isothermal hadronization hypersurface and
not for a general one hence it is specifically applicable to very high energy, where it is a better
approximation than the general formula, including temperature gradients (see discussion in \cite{Becattini:2021iol}). 
Second, the energy $\varepsilon = p \cdot \hat{t}=p^0$ is, in some studies, replaced by $p \cdot u$ 
where $u$ is the four-velocity field; this change leads to significant quantitative 
differences. Indeed, in Ref.~\cite{Alzhrani:2022dpi} the authors reproduce the correct sign of $P_z(\phi)$ without the isothermal decoupling but using $p\cdot u$, and this would not have been the case had they used $p^0$ \cite{chun}. We stress that the derivation adopted in Ref.~\cite{Becattini:2021suc} \emph{requires} to use $p^0$.}:
\begin{align}\label{eq:polarization isothermal}
    S^\mu(p) = - \epsilon^{\mu\rho\sigma\tau} p_\tau 
  \frac{\int_{\Sigma} \di \Sigma \cdot p \; n_F (1 -n_F) 
  \left[ \omega_{\rho\sigma} + 2\, \hat t_\rho \frac{p^\lambda}{\varepsilon} \Xi_{\lambda\sigma} \right]}
  { 8m T_{H} \int_{\Sigma} \di \Sigma \cdot p \; n_F},
\end{align}
where $T_H$ is the hadronization temperature, $\hat{t}^\mu=\delta^{\mu}_0$ is the time unit vector in
the QGP centre-of-mass frame, $\varepsilon= p \cdot \hat{t}$ and $n_F=[\e^{\beta\cdot p-\sum_i\mu_i/T}+1]^{-1}$ 
is the Fermi distribution (the chemical potentials can be taken as vanishing at the considered energies).
The presence of a specific time vector is a manifestation of the dependence of the above formula on the 
specific hadronization hypersurface, which in turn is related to the non-conservation of the local equilibrium
operator (see discussion in ref. \cite{Becattini:2021suc}).
The tensors $\omega_{\mu\nu}$ and $\Xi_{\mu\nu}$ are the kinematic vorticity and shear, respectively:
\begin{equation}\label{vortshear}
\omega_{\mu\nu}=\frac{1}{2}\left(\partial_\nu u_\mu-\partial_\mu u_\nu\right), \qquad \Xi_{\mu\nu}
=\frac{1}{2}\left(\partial_\nu u_\mu+\partial_\mu u_\nu\right).
\end{equation}
The formulae \eqref{eq:polarization isothermal} and \eqref{vortshear} make it apparent that, unlike 
most other hadronic observables, spin polarization is sensitive to the flow velocity gradients at 
leading order. Therefore, polarization can be used as an important observable to constrain various 
medium parameters, as we will show in the present study.

Over the past few years, there have been several $\Lambda$ polarization numerical studies with 
the hydrodynamic model of the QGP based on thermal vorticity \cite{Wu:2019eyi,Florkowski:2019voj,Huang:2020dtn,Lei:2021mvp,Serenone:2021zef,Singh:2021yba} or including 
also the shear tensor contribution 
\cite{Becattini:2021iol,Fu:2021pok,Florkowski:2021xvy,Alzhrani:2022dpi,Wu:2022mkr,Fu:2022myl,Jiang:2023fai,Jiang:2023vxp,Ribeiro:2023waz,Yi:2023tgg}. In this paper, we have carried out full numerical simulations of Au-Au 
at $\sqrts=200$ GeV and Pb-Pb collisions at $\sqrts=5020$ GeV and compared the obtained results to 
the available experimental data, employing up-to-date theoretical formulae for spin polarization. 
Furthermore, we have studied the sensitivity of spin polarization to the initial conditions
and transport coefficients, specifically the shear and bulk viscosity. For the first time, we 
have included in a realistic hydrodynamic simulations the corrections due to the decay of polarized 
$\Sigma^*$ and $\Sigma^0$ (the so-called feed-down corrections) to the local polarization. 
Additionally, we have studied the impact of varying the initial longitudinal momentum flow in the 
initial state, showing that the local polarization along the total angular momentum is quite sensitive 
to it. It should be pointed out that such a dependence was studied in great detail in 
ref. \cite{Jiang:2023vxp}, for Au-Au collisions at $\sqrts =27$ GeV. 
In order to avoid possible biases, we have employed two initial state models.

The paper is organized as follows. In section \ref{sec:numerics} we describe and validate the hydrodynamic
simulation setup using two different initial state models. Section \ref{sec:decays} is devoted to the 
study of feed-down corrections to the polarization vector, and in section \ref{sec:pol vs data}, we calculate various polarization observables measured experimentally. Finally, we address the effect of the initial collective flow and viscosity on polarization in 
section \ref{sec:visc}. 
\subsection*{Notation}
We adopt the natural units in this work, with $\hbar=c=K=1$. The Minkowskian metric tensor $g$ is ${\rm diag}(1,-1,-1,-1)$; for the Levi-Civita symbol we use the convention $\epsilon^{0123}=1$.
We will use the relativistic notation with repeated indices assumed to be summed over.

\section{Numerical framework}\label{sec:numerics}

The numerical setup used in this work builds upon the chain of codes \texttt{vHLLE} and \texttt{SMASH} \cite{Karpenko:2013wva,Schafer:2021csj,dmytro_oliinychenko_2023_7870822,SMASH:2016zqf}. 
Such a chain involves a pre-defined initial state, a 3+1D viscous hydrodynamic evolution of the produced 
dense medium, a fluid-to-particle transition (hadronization, or, more technically \emph{particlization}), taking 
place at a surface of fixed energy density, followed by a Monte-Carlo hadronic sampling from the particlization hypersurface and a subsequent hadronic rescattering and resonance decays.   

The hydrodynamic evolution is simulated with \texttt{vHLLE} code \cite{Karpenko:2013wva}, 
where the particlization hypersurface is reconstructed with \texttt{Cornelius} subroutine \cite{Huovinen:2012is}. 
This hypersurface is identified as the constant-energy-densty hypersurface $e=0.4$ GeV/fm$^3$, and it is used both for the evaluation of $\Lambda$ polarization as well as for conventional 
Monte-Carlo sampling of hadrons using \texttt{smash-hadron-sampler} code \cite{smash-hadron-sampler}. Finally, \texttt{SMASH} \cite{SMASH:2016zqf} handles rescatterings of the produced hadrons and decays 
of unstable resonances. We emphasize that spin degrees of freedom are not implemented in \texttt{SMASH} 
therefore, all the polarization results presented in this work (and so far in the literature) are 
solely based on formula \eqref{eq:polarization isothermal} 
and the subsequent calculation of polarization transfer in the decay of resonances. The polarization calculations are performed using a dedicated code, \texttt{hydro-foil}, which is publicly available 
\cite{hydro-foil}. To conveniently handle all the chain stages, we have created a hybrid 
model based on the Python programming language and Snakemake \cite{molder2021sustainable,repo}.  

\begin{table}
    \centering
    \begin{tabular}{c|c|c|c}
    \hline
    \hline
    Parameter & Description & @RHIC AuAu 200 GeV & @LHC PbPb 5020 GeV\\
    \hline
         $w[\text{fm}]$ &Size of the initial hot spot & 0.8& 0.14  \\
         $\eta_0$ &Size of the mid-rapidity plateau & 2.2& 2.7\\
         $\sigma_\eta$ &Space-time rapidity fall off width & 0.9& 1.2  \\
         $f$ &Initial longitudinal flow fraction & 0.15& 0.15 \\ 
         \hline
        \hline
    \end{tabular}
    \caption{Values of the free parameters in the initialization of the hydrodynamic stage using 
    \texttt{superMC}. For a detailed description of these initial condition parameters, see refs.~\cite{Shen:2020jwv,Alzhrani:2022dpi}.}
    \label{tab:parameters_supermc}
\end{table}
\begin{table}
    \centering
    \begin{tabular}{c|c|c|c}
    \hline
    \hline
    Parameter & Description & @RHIC AuAu 200 GeV & @LHC PbPb 5020 GeV\\
    \hline
         $R [\text{fm}]$ &Size of the initial hot spot & 0.4 & 0.4  \\
         $\eta_0$ &Size of the mid-rapidity plateau & 1.5 & 2.4\\
         $\sigma_\eta$ &Space-time rapidity fall off width & 1.4 & 1.4  \\
         $\eta_M$ & Initial state tilt parameter & 2.0 & 4.5  \\
         $\alpha$ & Fraction of binary scatterings & 0.15 & 0.15 \\
         \hline
        \hline
    \end{tabular}
    \caption{Values of the free parameters in the initialization of hydrodynamic 
    stage using 3D Glissando IS. For a detailed description of these initial condition 
    parameters, see refs.~\cite{Cimerman:2020iny}.}
    \label{tab:parameters_glissando}
\end{table}
\begin{figure}[h!]
    \centering
\includegraphics{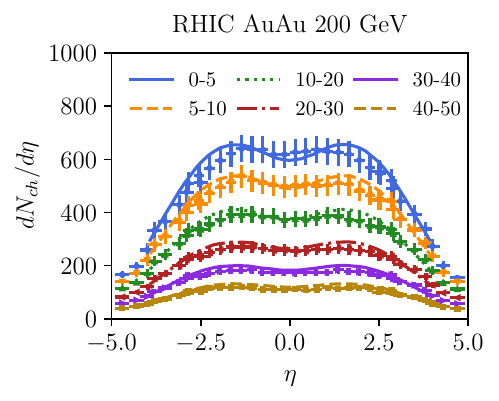}
\includegraphics{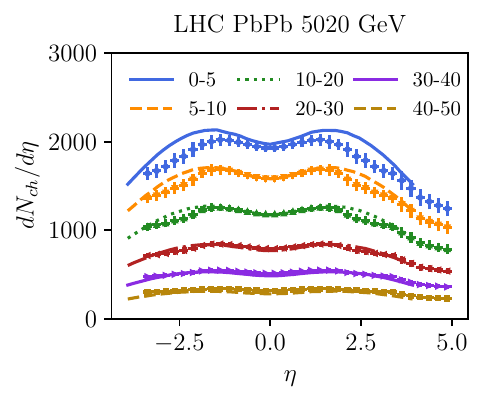}\\
\includegraphics{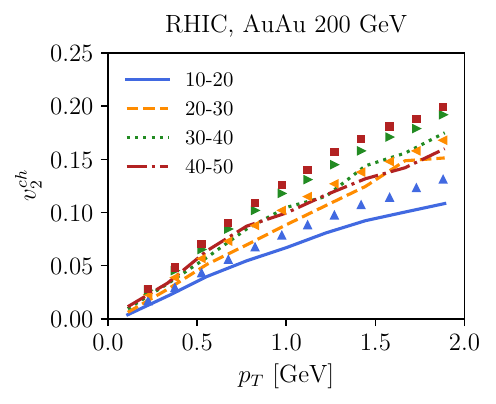}
\includegraphics{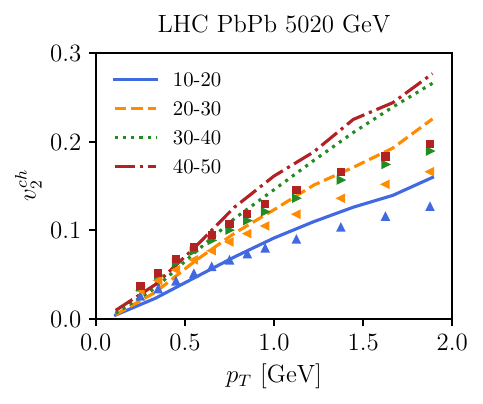}\\
\includegraphics{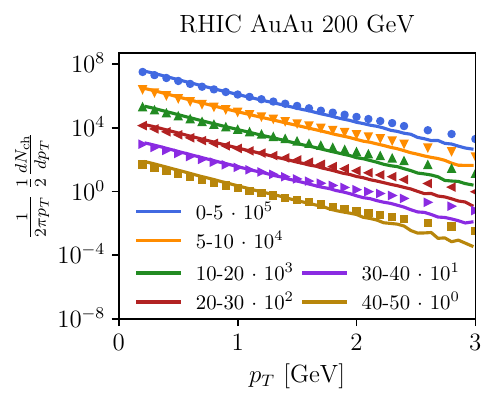}
\includegraphics{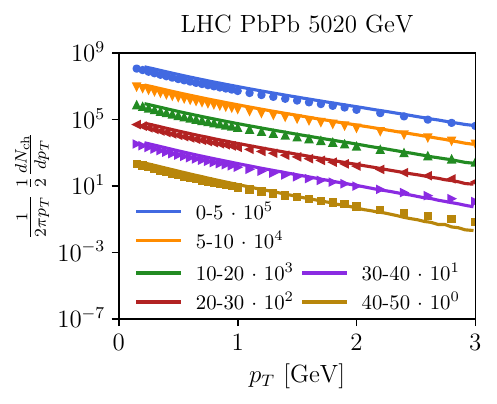}
\caption{Comparison between model and data with \texttt{superMC} initial conditions. 
Left panels correspond to Au-Au collisions at $\sqrts = 200$ GeV at RHIC and the 
right ones to $\sqrts = 5020$ GeV Pb-Pb collisions at LHC. In the upper panels, we show 
the pseudo-rapidity distribution of charged hadrons, with data from \cite{BRAHMS:2001llo} 
and \cite{ALICE:2016fbt}, the mid-panels show the elliptic flow of charged hadrons, 
data from refs. \cite{STAR:2004jwm} and \cite{ALICE:2022zks}, and the lower panels show the 
transverse momentum spectrum of charged hadrons, with data from refs. \cite{STAR:2003fka} 
and \cite{ALICE:2018vuu}. }
\label{fig:std_superMC}   
\end{figure}
\begin{figure}[h!]
    \centering
\includegraphics{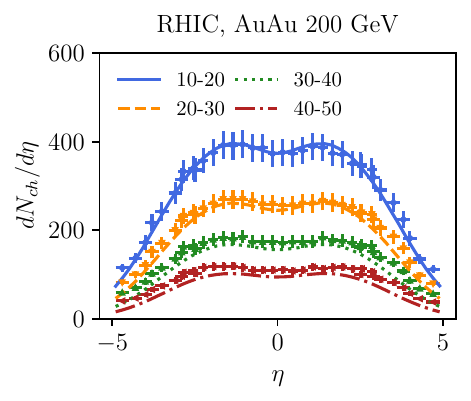}
\includegraphics{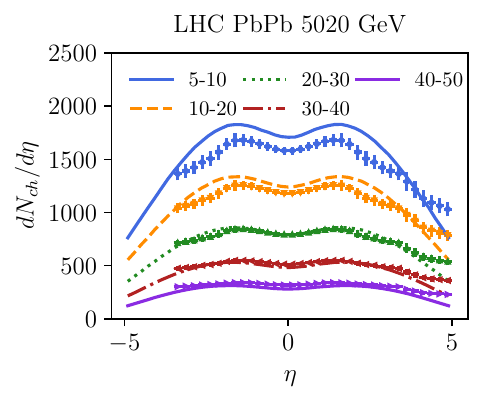}\\
\includegraphics{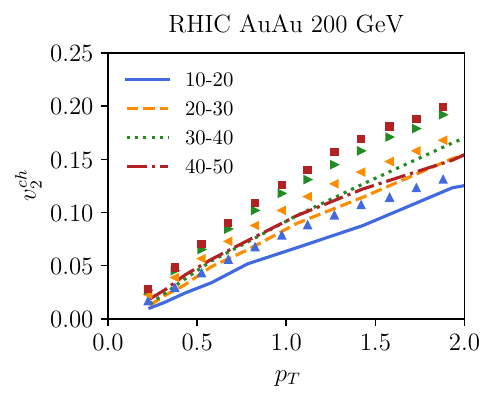}
\includegraphics{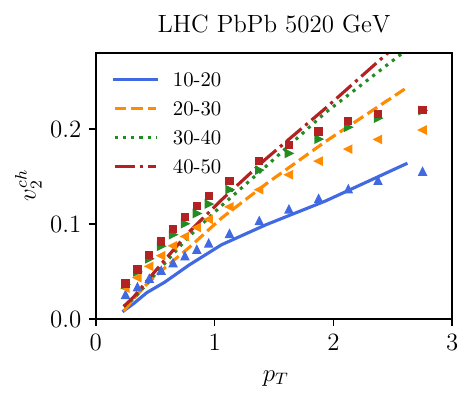}\\
\includegraphics{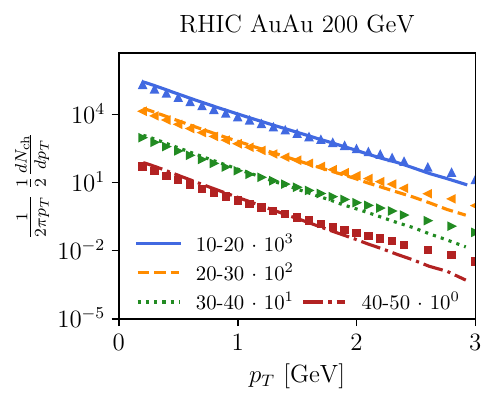}
\includegraphics{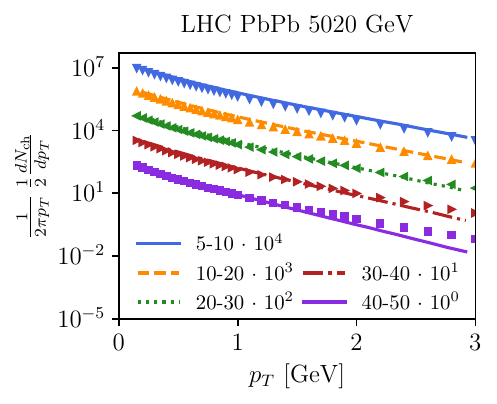}
    \caption{Comparison between model and experimental data with \texttt{GLISSANDO}
    initial conditions. Left panels correspond to Au-Au collisions at $\sqrts = 200$ GeV at RHIC 
    and the right ones to $\sqrts = 5020$ GeV Pb-Pb collisions at LHC. In the upper panels, we show the 
    pseudo-rapidity distribution of charged hadrons, the mid-panels show the elliptic 
    flow of charged hadrons, and the lower panels show the transverse momentum 
    spectrum of charged hadrons. The data are taken from the same references as in 
    fig. \ref{fig:std_superMC}.}
\label{fig:std_glis}
\end{figure}

As has been mentioned, we have used two Initial State (IS) models in this study. The first one is 
\texttt{superMC}, numerically implemented from scratch based on the formulae from \cite{Shen:2020jwv,Alzhrani:2022dpi}. 
\texttt{superMC} is based on Glauber geometry with local energy and momentum conservation conditions 
and provides a $\sqrt{T_A T_B}$ scaling of the energy density similar to TrENTo \cite{Moreland:2014oya} 
$p=0$ initial state. The second is a 3D extension of the initial state from Monte Carlo Glauber generator 
\texttt{GLISSANDO} \cite{Rybczynski:2013yba}, where entropy depositions from participant nucleons are taken to be tilted in space-time rapidity according to \cite{Bozek:2011ua}, whereas the depositions from binary scatterings are symmetric in space-time rapidity. With either IS option, we have used the starting time 
of the fluid stage $\tau_0$ and shear viscosity to entropy density ratio $\eta/s$ values that are considered 
optimal to reproduce basic hadronic observables. In the case of \texttt{superMC} IS, the 
starting time of the fluid stage is $\tau_0=1$ fm/c and, unless otherwise stated, a fixed shear viscosity 
over entropy density ratio of $\eta/s=0.08$. In the case of 3D \texttt{GLISSANDO} IS, the initial time 
is $\tau_0=0.4$~fm/c, $\eta/s=0.08$ at RHIC energy and $\eta/s=0.1$ at LHC energy, which reflects a 
somewhat higher temperature range probed in heavy-ion collisions at the LHC. Instead, for the bulk viscosity over 
entropy density $\zeta/s$, we have used the temperature-dependent parametrization introduced in \cite{Schenke:2020mbo}, referred to as 
parametrization III in section \ref{sec:visc} (see eq. \ref{eq:param3}).  

For $\sqrts=200$~GeV energy, the values of the free parameters of the \texttt{superMC} IS model 
have been slightly modified with respect to those originally published in \cite{Shen:2020jwv,Alzhrani:2022dpi} 
to obtain an optimal description of the pseudo-rapidity distribution, elliptic flow of charged hadrons, 
and transverse momentum spectrum of identified hadrons. For $\sqrts=5020$~GeV, a different 
optimal tuning has been found, with the values for both energies reported in Table~\ref{tab:parameters_supermc}. 
We have used an average initial state of 20k \texttt{superMC} events for the tuning. 
Figure \ref{fig:std_superMC} shows the results of our calculations compared to the experimental data. 
Baryon and electric charge currents do not play a significant role in collisions at the energies under 
consideration, so the free parameters associated with them, $\eta_{B0}$, $\sigma_{B\text{in}}$ and $\sigma_{B\text{out}}$, are the same as the ones reported in ref. \cite{Shen:2020jwv} for Au-Au 
collisions at $\sqrts=200$ GeV.

The 3D extension of \texttt{GLISSANDO} initial state is taken from the study in ref.\cite{Cimerman:2020iny}, and 
it follows a parametrization previously reported in \cite{Bozek:2015bha}. Differently from 
ref. \cite{Cimerman:2020iny}, we have not computed the normalization of the initial state energy density 
by identifying the total energy of the fluid with the sum of energies of incoming participants. Instead, 
we have treated normalization as another free parameter and tuned it to fit charged hadron $dN/d\eta$
distribution. The resulting parameters are reported in Table~\ref{tab:parameters_glissando}, and 
figure~\ref{fig:std_glis} shows the results of our calculations. We generally have a very good agreement 
of transverse momentum spectra and pseudo-rapidity distributions of charged hadrons at various centralities but a less good agreement for their elliptic flow coefficient. We attribute the underestimated elliptic flow signal with both initial states to fluid dynamic description with averaged initial state, as compared to event-by-event fluid-dynamic picture in other studies. We also note that at the LHC energy the agreement with the data is better, as the ensuing fluid stage has a longer lifetime, therefore the final flow is somewhat less sensitive to the presence of fluctuations in the initial stage.

\section{Feed-down contribution to polarization}\label{sec:decays}

The hydrodynamic-statistical model predicts that all particles created by the particlization
of the QGP are polarized. When unstable particles are produced in this stage, they transfer part of their polarization to their decay products. Therefore, secondary $\Lambda$s produced by decaying heavier particles or resonances are also polarized. The contribution to polarization owing to the secondary $\Lambda$s is called feed-down
correction. As the experiments can subtract $\Lambda$s from long-lived weakly decaying 
hadrons, we only consider short-lived hadrons decaying through electromagnetic or strong interaction. 
Specifically we focus on $\Sigma^*\rightarrow\Lambda +\pi$ and $\Sigma^0\rightarrow \Lambda + \gamma$,
which provide the predominant channels to secondary $\Lambda$ production.

\begin{figure}[h!]
    \centering
    \includegraphics{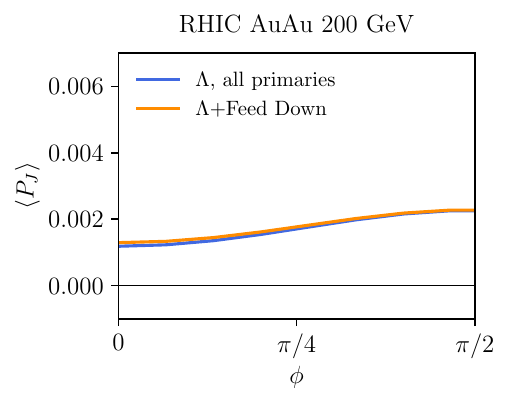}
    \includegraphics{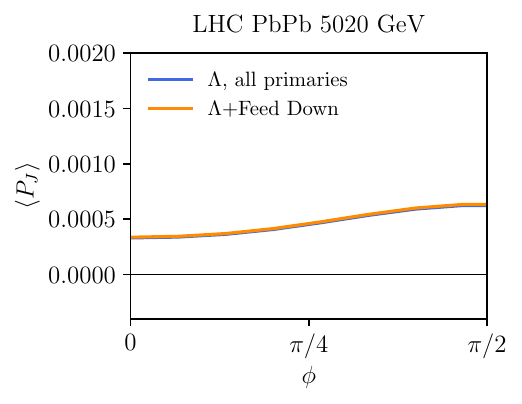}\\
    \includegraphics{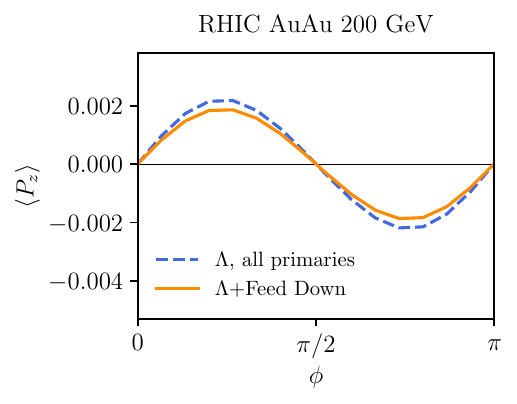}
    \includegraphics{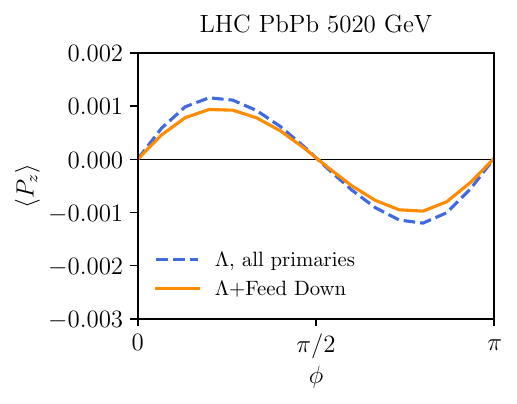}
    \caption{Feed-down corrections on the polarization along the angular momentum 
    direction (upper panels) and along the beam line (lower panels). The solid line shows the
    polarization including the contribution of $\Sigma^*$ and $\Sigma^0$ while the 
    dashed line the calculated polarization assuming that all $\Lambda$s are primaries.
    The simulations are in the centrality window 20-60\% for RHIC and 30-50\% for LHC.}
    \label{fig:feed-down}
\end{figure}

The polarization transferred from a particle to its decay products in a two-body decay, 
has been extensively studied elsewhere \cite{Becattini:2016gvu,Xia:2019fjf,Becattini:2019ntv};
herein, we briefly review it and describe the formulae used to compute it. 
Let us then consider the general decay $M\rightarrow \Lambda D$. The particle $M$ produced at hadronization is polarized, and, if it is a fermion, its mean 
spin vector can be obtained by simply rescaling the equation \eqref{eq:polarization isothermal} \cite{Becattini:2016gvu,Xia:2019fjf,Becattini:2019ntv,Palermo:2023cup}:
\begin{align}\label{eq:polarization isothermal generic S}
   S^\mu(p) = - \frac{S(S+1)}{3}\epsilon^{\mu\rho\sigma\tau} p_\tau 
  \frac{\int_{\Sigma} \di \Sigma \cdot p \; n_F (1 -n_F) 
  \left[ \omega_{\rho\sigma} + 2\, \hat t_\rho \frac{p^\lambda}{\varepsilon} \Xi_{\lambda\sigma} \right]}
  { 2m T_{H} \int_{\Sigma} \di \Sigma \cdot p \; n_F},
\end{align}
where $S$ is the spin of the fermion. As it has been discussed, the produced $\Lambda$ is polarized.
The spin vector ${\bf S}_*^{(M)}$ of the $\Lambda$ particle in its rest frame\footnote{The spin four-vector in the rest frame is $S^{\mu}= (0 ,{\bf S}_*^{(M)} )$.} inherited from the mother particle is \cite{Becattini:2019ntv}:
\begin{align}\label{eq:inherited spin}
    {\bf S}_*^{(M)}({\bf p})=\frac{\int \di \Omega_* n({\bf P})F({\bf p},\Omega_*) {\bf S}_{M\to\Lambda}
    ({\bf P},{\bf p})}{\int \di \Omega_* n({\bf P})F({\bf p},\Omega_*)}.
\end{align}
In the above formula, the solid angle $\Omega_*$ is in the rest frame of the mother particle, 
${\bf p}$ and ${\bf P}$ are the momenta of $\Lambda$ and $M$ respectively in the QGP frame and ${\bf S}_{M \to \Lambda}$ 
is a ``spin-transfer'' vector function depending on the decay. For the aforementioned decays, they read:
\begin{subequations}\label{subeq: S vectors}
\begin{align}
{\bf S}_{\Sigma^*\to \Lambda}({\bf P},{\bf p}) & =\frac{2}{5}\left({\bf S}_{*\Sigma^*}({\bf P})-
\frac{1}{2}\hat{{\bf p}}\cdot {\bf S}_{* \Sigma^*}({\bf P}) \;\hat{{\bf p}}\right)\\
{\bf S}_{\Sigma^0\to\Lambda}({\bf P},{\bf p}) &=-\hat{{\bf p}} \cdot {\bf S}_{* \Sigma^0}({\bf P}) 
\;\hat{{\bf p}}
\end{align}
\end{subequations}
where ${\bf S}_{* \Sigma^*}$ and ${\bf S}_{* \Sigma^0}$ are the spin vectors of the mother particles 
in the mother's rest frame, which can be computed with eq. \eqref{eq:polarization isothermal generic S}.
The function $F({\bf p},\Omega_*)$ is given by:
\begin{align}
 F({\bf p},\Omega_*)=\frac{m_\Lambda^3(\varepsilon_{*\Lambda}+\varepsilon_\Lambda)
 \left[(\varepsilon_{*\Lambda}+\varepsilon_\Lambda)^2-(\varepsilon_\Lambda\varepsilon_{*\Lambda}+{\bf p}
 \cdot{\bf p}_*+m_\Lambda^2)\right]}{\varepsilon_{* \Lambda}(\varepsilon_\Lambda\varepsilon_{*\Lambda}+
 {\bf p}\cdot{\bf p}_*+m_\Lambda^2)^3},
\end{align}
where $\varepsilon_{* \Lambda}$ and ${\bf p}_*$ are the energy and the momentum of the $\Lambda$ particle 
in the mother's rest frame. Finally, the function $n({\bf P})$ is the momentum spectrum of $M$ in the
QGP frame obtained from the Cooper-Frye prescription:
\begin{align}
    n({\bf P})=\frac{\di N}{\di^3 {\rm P}}=\frac{1}{\varepsilon_P}\int\di\Sigma\cdot P \; 
    \frac{1}{\e^{\beta\cdot P-\mu/T}+1}.
\end{align}
All the integrals in equation \eqref{eq:inherited spin} are evaluated in the \texttt{hydro-foil} code.

To determine the overall $\Lambda$'s polarization, one should know the fraction of primary $\Lambda$s
and those from the considered decays. Although such fractions are, in principle, momentum dependent, 
here, for simplicity, we assume that they are constant and use the numbers quoted in ref. 
\cite{Becattini:2019ntv} estimated by using the statistical hadronization model: the fraction of primary 
$\Lambda$, at very high energy, is $f_P=0.243$ whereas for the secondary $\Lambda$ one has respectively $f_{\Sigma^*}=0.359$ and $f_{\Sigma^0} = 0.275\times 0.6 =0.165 $, where the $0.6$ takes into account that 
only $\approx60\%$ of the $\Sigma_0$ are primary; other sources are neglected. Finally, the total $\Lambda$'s spin 
vector in its rest frame is calculated by normalizing it accordingly:
\begin{align}\label{eq:tot spin decay}
    {\bf S}_{*}^{\text{tot}}(p)=\frac{f_P {\bf S}_{*}^{\text{P}}({\bf p})+f_{\Sigma^*} 
    {\bf S}_*^{(\Sigma^*)}({\bf p})+f_{\Sigma^0} {\bf S}_*^{(\Sigma^0)}({\bf p})}{f_P+
    f_{\Sigma^*}+f_{\Sigma^0}},
\end{align}
where ${\bf S}_{*}^{\text{P}}({\bf p})$ is the spin polarization vector of primary $\Lambda$s, 
determined by boosting the spin vector in the eq. \eqref{eq:polarization isothermal} back to the 
$\Lambda$ rest frame and ${\bf S}_*^{(\Sigma^*)}({\bf p})$, ${\bf S}_*^{(\Sigma_0)}({\bf p})$ are 
calculated by means of the eq. \eqref{eq:inherited spin}.

Figure \ref{fig:feed-down} shows the effect of the feed-down contribution to the component of
the polarization vector along the angular momentum (henceforth referred to as {\em transverse})
and along the beamline (henceforth referred to as longitudinal) $\langle P_{J,z} \rangle
= 2 \langle S_{J,z} \rangle$ by comparing it to a calculation where all $\Lambda$ are assumed to be 
primary. The simulations are performed with the \texttt{superMC} initial conditions. 
As already remarked in \cite{Xia:2019fjf,Becattini:2019ntv} the combination in \eqref{eq:tot spin decay} 
results in an accidental approximate cancellation so that the total polarization is almost 
the same as though the $\Lambda$s were only primary. More precisely, the feed-down correction 
implies a $\approx 10\%$ reduction to the full primary case for the longitudinal 
component and $\approx 3\%$ for the transverse component. So, even with the contribution
from the thermal shear tensor included, which was not the case in refs. \cite{Xia:2019fjf,Becattini:2019ntv},
we confirm the previous conclusion that adding feed-down corrections does not significantly change the 
polarization with respect to the simple assumption of an entirely primary 
production. 

\section{Polarization at very high energy: results}\label{sec:pol vs data}

We will now present the results of our calculation and their comparison with the data.
In our calculations, feed-down corrections are always included; even though they are small
corrections to the assumption of pure primary production, they certainly improve the accuracy 
of theoretical predictions. In the plots, we show the proper polarization vector $P^\mu(p)=S_*^\mu(p)/S$ 
where $S$ is the spin of the particles and $S_*^\mu(p)$ is the mean spin vector in the $\Lambda$'s rest 
frame (hence, for $\Lambda$ hyperon $P^\mu=2 S_*^\mu$). The value of the $\Lambda$ decay constant 
has been set to $\alpha_\Lambda=0.732$ to compare with experiments \cite{ParticleDataGroup:2020ssz}.

Simulations have been carried out using averaged initial states, which, as has been discussed 
in section \ref{sec:numerics}, were defined by the models \texttt{superMC} and \texttt{GLISSANDO}. For the \texttt{superMC} IS, the initial state was generated by averaging 20k samples. For \texttt{GLISSANDO} IS, 
2k -- 15k samples were used, depending on centrality. We conduct the study with event-averaged initial states, to show generic effects of bulk viscosity on polarization observables. We note that, in an event-by-event fluid dynamic description, a fluctuating and spiky initial state will result in a stronger radial flow, which will need to be compensated by a larger bulk viscosity in order for the mean $p_T$ to agree with an experimentally measured value. Therefore in a more realistic event-by-event simulation, we expect the effects of bulk viscosity to be even larger than in the present study.

In all cases, we compute polarization of $\Lambda$s 
in the rapidity window $\eta\in[-1,1]$ to match experimental measurements. Polarization as 
a function of the azimuthal angle $\langle P_z (\phi) \rangle$ and $\langle P_J(\phi) \rangle$ are 
integrated over the $p_T$ range $(0,6)$ GeV weighted by the $\Lambda$ spectrum, whereas 
$\langle P_z \sin 2\phi \rangle$ is the mean over the azimuthal angle $\phi$. 

\begin{figure}[h!]
    \centering
\includegraphics{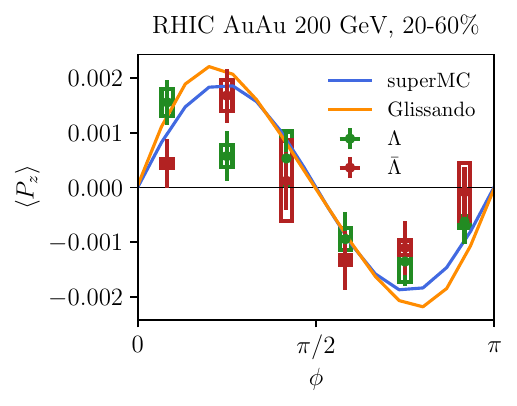}
\includegraphics{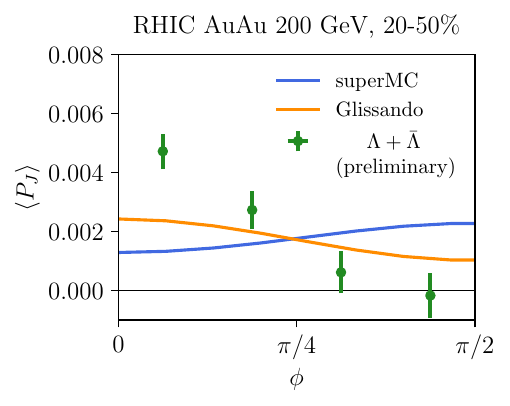}\\
\includegraphics{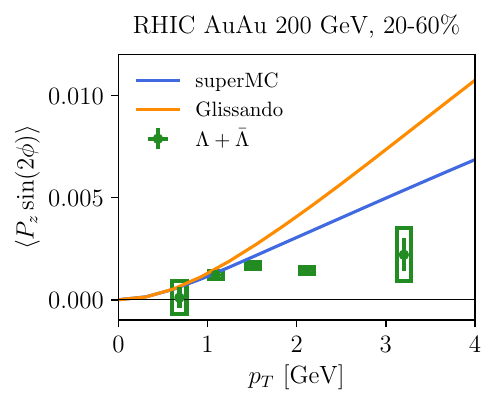}
\includegraphics{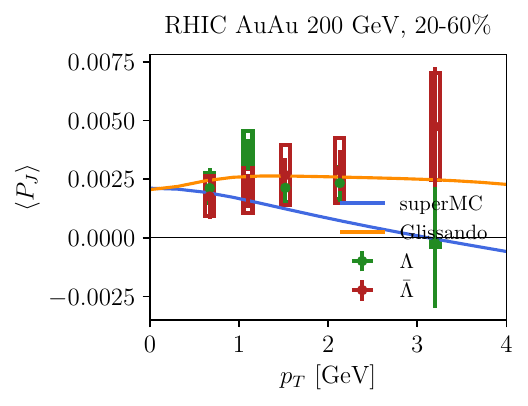}\\
\includegraphics{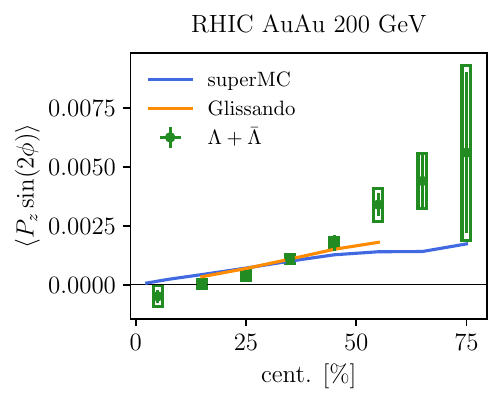}
\includegraphics{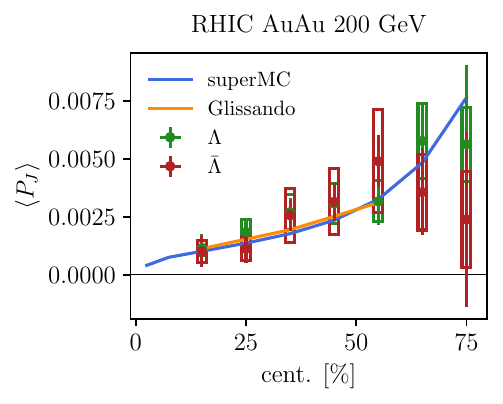}
    \caption{Comparison between theoretical calculations and experimental data for Au-Au 
    collisions at 200 GeV. Top panels: azimuthal angle dependence of longitudinal and transverse
    polarization. The middle panels show the transverse momentum dependence of $\langle P_z\sin(2\phi) \rangle$
    and $\langle P_J \rangle$, and the bottom panels show their dependence on centrality. For the top and middle panels, the simulations are carried 
    in the centrality 20-60\%. The data are taken from refs.
    \cite{Niida:2018hfw,STAR:2018gyt,STAR:2019erd}.}
    \label{fig:data_rhic}
\end{figure}

\begin{figure}[h!]
    \centering
\includegraphics{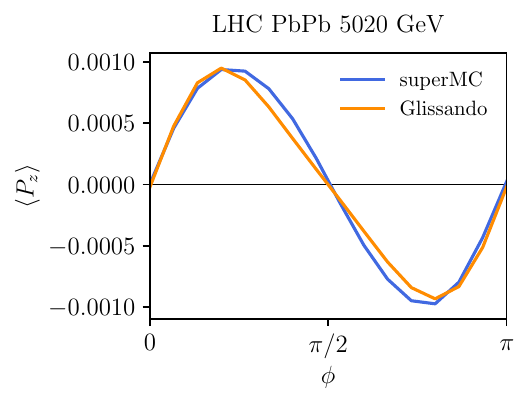}
\includegraphics{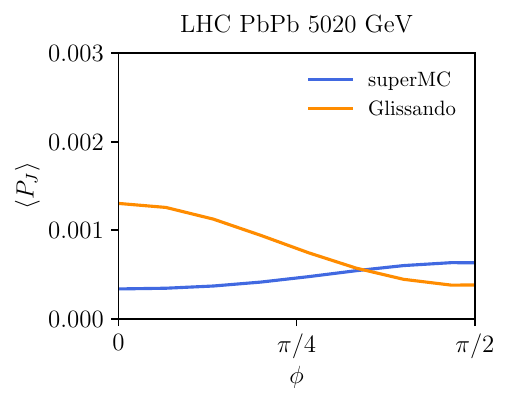}\\
\includegraphics{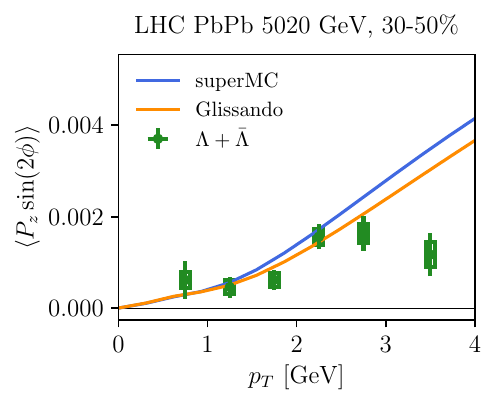}
\includegraphics{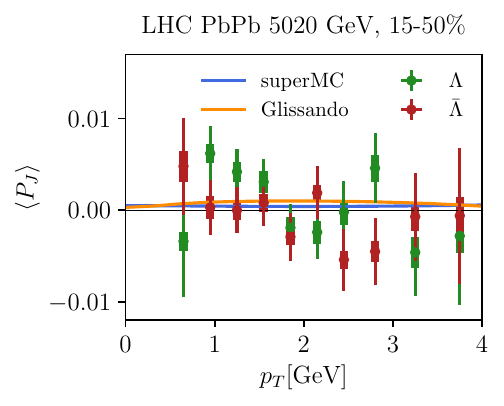}\\
\includegraphics{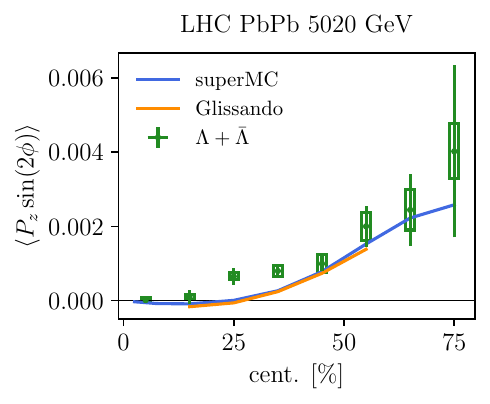}
\includegraphics{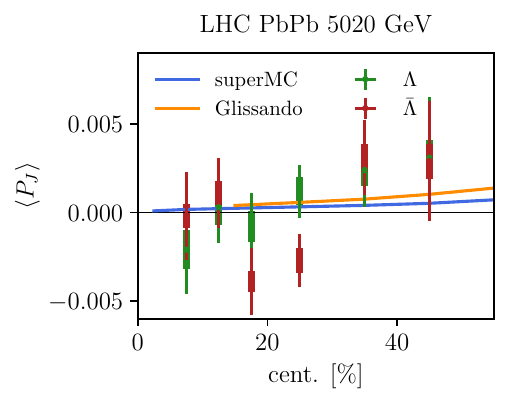}
    \caption{Comparison between theoretical calculations and experimental data for Pb-Pb 
    collisions at 5020 GeV. Top panels: azimuthal angle dependence of transverse and longitudinal
    polarization. The middle panels show the transverse momentum dependence of $\langle P_z\sin(2\phi) \rangle$
    and $\langle P_J \rangle$, and the bottom panels show their dependence on centrality. For the top and middle panels, the simulations are carried 
    in the centrality 30-50\%. The data are taken from refs.
     \cite{ALICE:2019onw,ALICE:2021pzu}
    .}
    \label{fig:data_lhc}
\end{figure}

For $\sqrts = 200$ Gev Au-Au collisions, the results are shown in figure \ref{fig:data_rhic}
where the data points have been taken from refs. \cite{Niida:2018hfw,STAR:2018gyt,STAR:2019erd}. The centrality reported in the titles in the top and middle panels refers to the data, whereas simulations have always been carried out at 20-60\%. 
Our model describes well the longitudinal component of polarization $P_z$, as can 
be seen in the figures, although we overshoot the data in the high $p_T$ region. This result 
confirms the previous finding \cite{Becattini:2021iol} that the formula \eqref{eq:polarization isothermal} 
reproduces the data, essentially solving the so-called polarization sign puzzle. 
The difference between our calculations and those in ref. \cite{Alzhrani:2022dpi} 
using the same initial conditions \texttt{superMC} is owing to the use of the isothermal hadronization
assumption as well as a difference in the expression of energy $\varepsilon$ in \eqref{eq:polarization isothermal}, which is $p \cdot \hat t$ in our case and $p \cdot u$ in ref. \cite{Alzhrani:2022dpi}.

While \texttt{GLISSANDO} and \texttt{superMC} provide similar predictions for the longitudinal
polarization, they differ for the transverse component and especially for its azimuthal angle dependence. \texttt{GLISSANDO} IS predicts the correct dependence of $\langle P_J\rangle$ on the azimuthal
angle, although with an overestimated polarization. Instead, \texttt{superMC} does not get the slope right, although the 
global polarization (middle-right panel) is well reproduced; similar observations were reported in \cite{Alzhrani:2022dpi}
which, as has been mentioned, used the same IS model. In general, as it can be seen in the other 
plots in figure \ref{fig:data_rhic} \texttt{GLISSANDO} predicts a larger $P_J$ than \texttt{superMC} 
as a function of $p_T$ and centrality.

A substantial difference in the implementation of {\em longitudinal} 
hydrodynamic initial conditions causes the discrepancy between \texttt{GLISSANDO} and \texttt{superMC} in the transverse component $P_J$. Indeed, it should be stressed that the transverse component of the polarization is sensitive to the longitudinal velocity (see the equation \eqref{eq:polarization isothermal}) whereas the longitudinal component of 
the polarization is sensitive to the transverse velocity \cite{Becattini:2017gcx,Voloshin:2017kqp,Jiang:2023vxp}. While 
the hydrodynamic initial conditions in the transverse plane are strongly constrained by the observed 
transverse momentum spectra and their azimuthal anisotropies, initial longitudinal flow conditions 
are not accurately known. Indeed, \texttt{superMC} IS parametrization entails a non-vanishing $\tau$-$\eta$ 
component of the energy-momentum tensor, hence a non-vanishing initial longitudinal velocity, 
$u^\eta\neq 0$, whereas \texttt{GLISSANDO} IS initializes $u^\eta=0$. This difference will be 
addressed in more detail in the next section.

The results of our calculations for Pb-Pb collisions at $\sqrts = 5020$ GeV are shown in 
figure \ref{fig:data_lhc} along with the ALICE data \cite{ALICE:2019onw,ALICE:2021pzu}. The top panels show the theoretical predictions of longitudinal and transverse polarization as a function of the azimuthal angle: no data is available for these observables yet. In the middle panels, once again, the centrality reported in the title refers to the data, whereas simulations have been carried out at 30-50\%. The predicted 
longitudinal polarization is smaller than at RHIC energy by a factor of $\approx 2$ with both IS conditions. 
It can be seen that our calculations for 
the longitudinal polarization with both IS models agree with each other, as well as with the data (when available). On the other hand, 
like at RHIC energy, the predictions of the two IS models used are very different for the transverse polarization. The measurement of $P_J$ at this energy has significantly larger error bars than RHIC 
energy, so no definite conclusion can be drawn from comparing the predictions with the data.

\begin{figure}
    \centering
\includegraphics{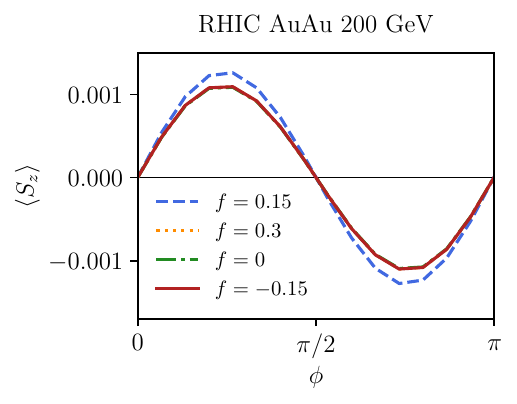}
\includegraphics{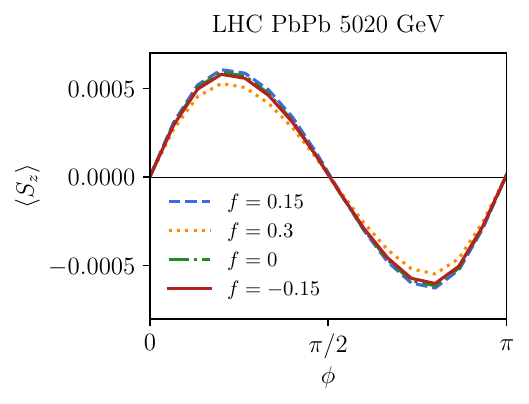}\\
\includegraphics{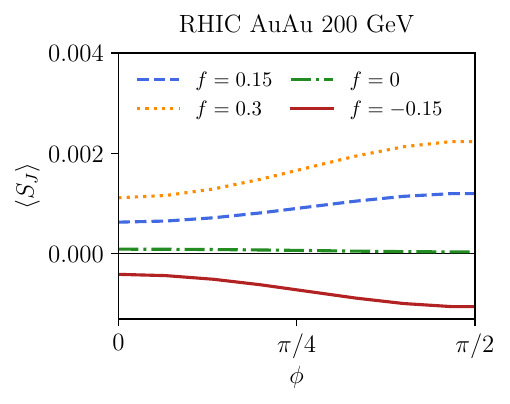}
\includegraphics{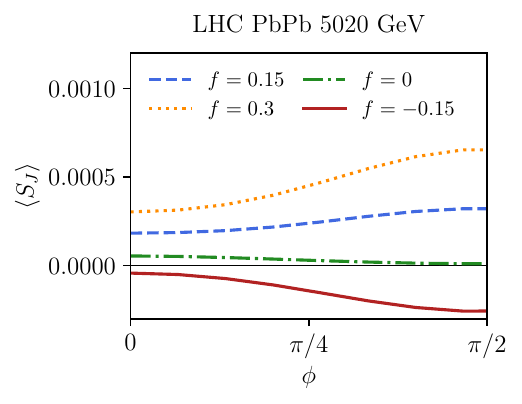}
    \caption{Azimuthal angle dependence of $S_z$ (upper panels) and $S_J$ (lower panels) for 
    different values of the parameter $f$ in the \texttt{superMC} initial state model. RHIC simulations are made in 20-60\% centrality, whereas LHC at 30-50\%.} 
    \label{fig:pol vs f}
\end{figure}

\section{Sensitivity to initial conditions and transport coefficients}\label{sec:visc}

In this section, we delve into the capability of spin polarization as a probe of the hydrodynamic
initial conditions and the transport coefficients of the QGP. We already mentioned that the two
IS used differ in the predictions of the transverse polarization $P_J$, which reflects different initial conditions for the longitudinal momentum density. To make this apparent, we have 
studied the sensitivity of $P_z(\phi)$ and $P_J(\phi)$ to the parameter $f$ in the \texttt{superMC} 
IS model. This parameter determines the initial values of the components of the energy-momentum 
tensor in the Milne coordinates as follows:
\begin{subequations}\label{eq:initial cond}
    \begin{align}
    T^{\tau\tau}&=\rho \cosh(f\,y_{CM})\\
    T^{\tau\eta}&=\frac{\rho}{\tau}\sinh(f\,y_{CM})
    \end{align}
\end{subequations}

where $\rho$ is the local energy density distribution and $y_{CM}$ is the center of mass rapidity 
(for a full description of these quantities, see ref. \cite{Alzhrani:2022dpi}). Therefore, the value 
of $f$ characterizes the initial longitudinal momentum density $T^{\tau\eta}$, thus driving the initial longitudinal flow and the fireball's total angular momentum. The latter affects the transverse component of the polarization $P_J$ as already discussed. We note that the equations \eqref{eq:initial cond} have
been considered also in ref. \cite{Jiang:2023vxp}, where the dependence on $f$ of the $\Lambda$ 
polarization has been studied at $\sqrts= 27$ GeV.
Figure \ref{fig:pol vs f} shows both transverse and longitudinal polarization from hydrodynamic simulations of Au-Au and Pb-Pb collisions initialized with different values of $f$. As expected, 
while the longitudinal component is almost insensitive to $f$, the transverse component changes 
significantly in magnitude, slope, and even sign. This sensitivity makes it possible to use 
local transverse polarization to discriminate between different models of initial longitudinal 
conditions and to pin down the amount of initial longitudinal momentum density. For the case at hand,
the curves are shown in figure \ref{fig:pol vs f}, and the results from \texttt{GLISSANDO} IS 
presented in the section \ref{sec:pol vs data} seem to favor the option $u^\eta=0$, implemented 
by $f=0$, which is the only one yielding at least the correct sign of the slope of the function 
$P_J(\phi)$ at RHIC.

\begin{figure}
    \centering
    \includegraphics{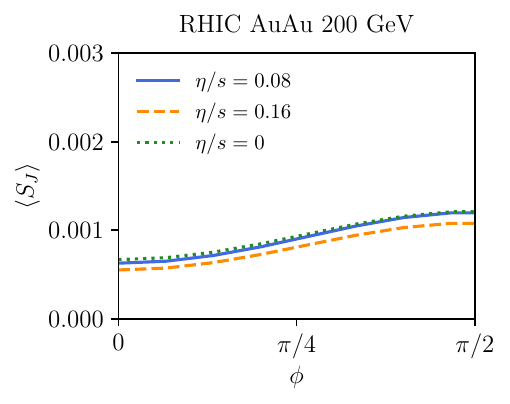}
\includegraphics{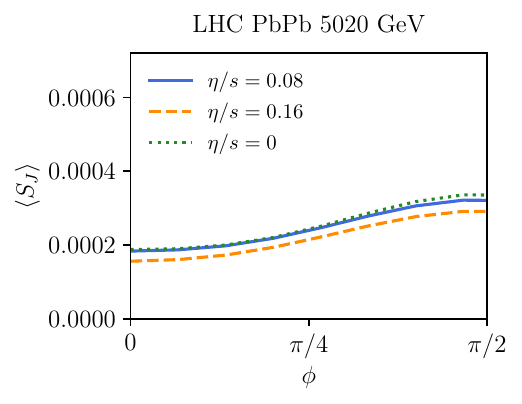}\\
\includegraphics{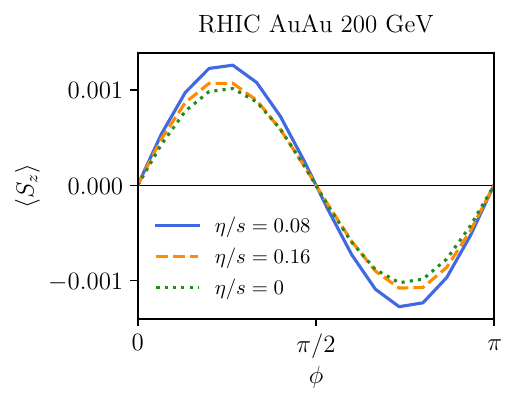}
\includegraphics{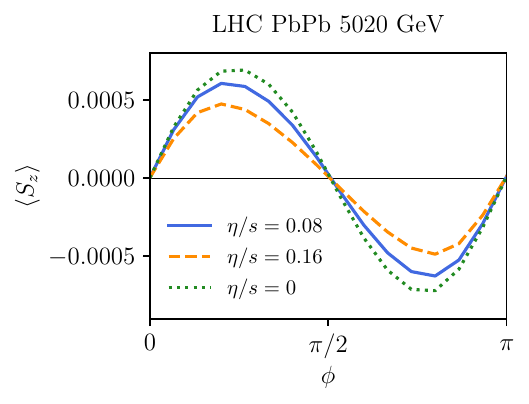}
    \caption{Transverse and longitudinal components of the spin vector as a function of the 
    azimuthal angle $\phi$ for various values of the shear viscosity over entropy density. 
    The centrality window is 20-60\% for RHIC and 30-50\% for LHC. }
    \label{fig:etas}
\end{figure}
We have also studied the dependence of the polarization on the shear viscosity and the bulk viscosity.
In figure \ref{fig:etas}, the transverse and longitudinal components of the spin vector of 
primary $\Lambda$s with the \texttt{superMC} IS, averaged over the same kinematic intervals as in section \ref{sec:pol vs data}, are shown at RHIC and LHC energies for different constant ratios of $\eta/s$ whereas the bulk viscosity is given by the parametrization III as explained in sec. \ref{sec:numerics}. 
The transverse component is almost insensitive to the shear viscosity, while the longitudinal 
component has a limited sensitivity. Interestingly, we notice that while at RHIC energies, a larger 
viscosity enhances the signal, at LHC, it slightly reduces it.

\begin{figure}
    \centering
\includegraphics{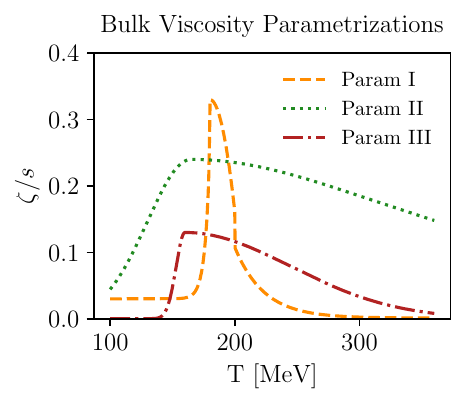}
    \caption{The temperature dependence of the bulk viscosity over entropy density ratio is used in this work.}
    \label{fig:bulks}
\end{figure}

\begin{figure}
    \centering
    \includegraphics{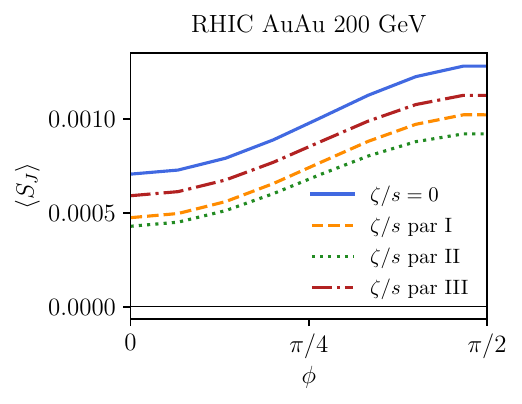}
\includegraphics{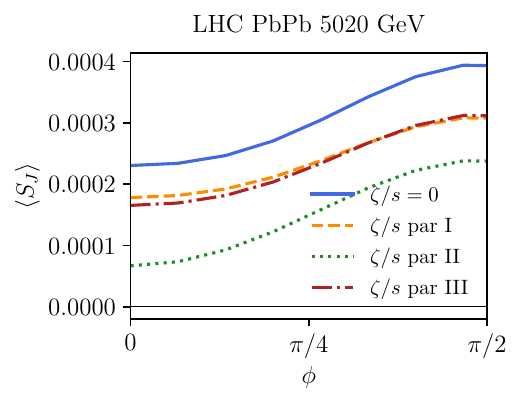}\\
\includegraphics{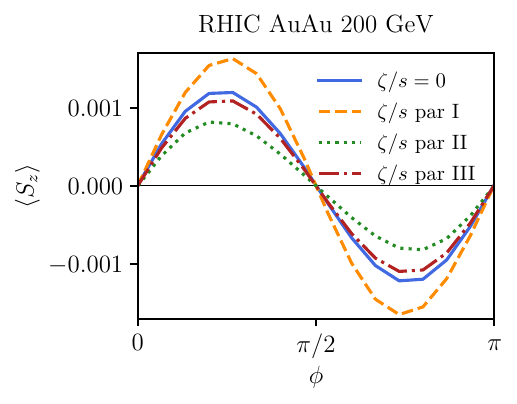}
\includegraphics{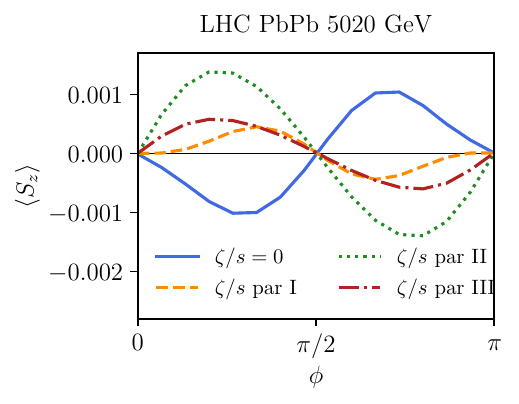}
    \caption{Transverse and longitudinal components of the spin vector as a function of 
    the azimuthal angle $\phi$ for various bulk viscosity parametrization at $\sqrts = 200$ 
    GeV (20-60\% centrality) and $\sqrts = 5020$  GeV (30-50\% centrality). For these plots we used the superMC initial state.}
    \label{fig: bulk params}
\end{figure}

Conversely, bulk viscosity has a sizeable effect on polarization, especially at high energy,
in agreement with previous observations \cite{Palermo:2022lvh}. We have used three different 
parametrizations of bulk viscosity as a function of temperature \cite{Bobek:2022xks}. The 
first one, dubbed as ``Parametrization I'' was introduced in ref. \cite{Ryu:2017qzn}:
\begin{equation}\label{eq:param1}
\zeta / s=\begin{cases}
c_1+0.08 \exp\left[\frac{T/T_{p}-1}{0.0025}\right]+
0.22\exp\left[\frac{T / T_{p}-1}{0.0022}\right]
& \quad T<T_p\\
c_2+27.55\left(T / T_{p}\right)-
13.77\left(T / T_{p}\right)^{2} 
& \quad T_p<T<T_P\\
c_3+0.9 \exp\left[\frac{-\left(T / T_{p}-1\right)}{0.0025}\right]+0.25 \exp\left[\frac{-\left(T / T_{p}-1\right)}{0.13}\right] 
& \quad T>T_P,
\end{cases}
\end{equation}
where $T_p=180$~MeV, $T_P=200$~MeV, $c_1=0.03$, $c_2=-13.45$ and $c_3=0.001$. The second 
parametrization was introduced in ref.~\cite{Schenke:2019ruo}:
\begin{equation}\label{eq:param2}
\zeta / s=\begin{cases}
B_{\mathrm{norm}}\exp\left[-\left(\frac{T-T_{\mathrm{peak}}}{T_{\mathrm{width}}}\right)^2\right]
& T<T_{\mathrm{peak}}
\\
B_{\mathrm{norm}}\frac{B_{\mathrm{width}}^2}{(T/T_{\mathrm{peak}}-1)^2+B_{\mathrm{width}}^2}
& T>T_{\mathrm{peak}},
\end{cases}
\end{equation}
where $T_{\mathrm{peak}}=165$~MeV, $B_{\mathrm{norm}}=0.24$, $B_{\mathrm{width}}=1.5$ and 
$T_{\mathrm{width}}=50$~MeV. The last parametrization, dubbed as ``Parametrization III'', was 
introduced in ref. \cite{Schenke:2020mbo} and it reads:
\begin{equation}\label{eq:param3}
\zeta / s=\begin{cases}
B_{\rm n}\exp\left[-\left(\frac{T-T_{\mu}}{B_1}\right)^2\right]
& T<T_{\mu}
\\
B_{\rm n}\exp\left[-\left(\frac{T-T_{\mu}}{B_2}\right)^2\right]
& T>T_{\mu},
\end{cases}
\end{equation}
where $B_{\rm n}=0.13$, $B_1=10$~MeV, $B_2=120$~MeV and $T_{\mu}=160$~MeV. 
These parametrizations are shown in figure \ref{fig:bulks}, where it can be seen that parametrization I 
has the sharpest peak around the transition temperature, the other two featuring a broader peak. In 
parametrization II, $\zeta/s$ is always larger than parametrization III.\\
The components of the mean spin vector - of primary particles only - along the angular momentum direction 
and the beam axis are shown in figure \ref{fig: bulk params} in Au-Au collisions at $\sqrts = 200$ GeV and Pb-Pb 
collisions at $\sqrts = 5020$ GeV for all the different parametrizations and using the \texttt{superMC} IS (with $f=0.15$ as in the initial tuning).
Bulk viscosity's effect on polarization is mild at $\sqrts = 200$ GeV: different parametrizations slightly 
change the magnitude of the polarization signal compared to the case of $\zeta=0$, but their 
difference is well within the experimental errors. Instead, surprisingly, at $\sqrts = 5020$ GeV the effect 
of bulk viscosity is dramatic: the longitudinal polarization flips/changes sign when bulk viscosity is 
introduced, and the difference between the three parametrizations is large. Identifying a  
single effect responsible for such different behavior at RHIC and LHC energies is hard. 
A possible explanation could be that
the higher average temperature and the longer lifetime of the QGP at the LHC energy
make the role of bulk viscosity more important.\\
To investigate this phenomenon in more detail, we have computed the contribution of different 
hydrodynamic gradients to the spin vector for primary $\Lambda$s. Referring to eq. \eqref{eq:polarization isothermal}, 
the kinematic vorticity and shear can be decomposed in terms of acceleration $A^\mu=u\cdot \partial u^\mu$, 
angular velocity $\omega^\mu=-1/2 \,\epsilon^{\mu\nu\rho\sigma}\partial_\rho u_\nu u_\sigma$, shear 
stress tensor $\sigma^{\mu\nu}$ and expansion scalar $\theta$ so that:
\begin{align}
\omega_{\mu\nu}&=\epsilon_{\mu\nu\rho\sigma}\omega^\rho u^\sigma+\frac{1}{2}\left(A_\mu u_\nu-A_\nu u_\mu\right),\\
    \Xi_{\mu\nu}&=\frac{1}{2}\left(A_\mu u_\nu+A_\nu u_\mu\right)+\sigma_{\mu\nu}+\frac{1}{3}\theta \Delta_{\mu\nu},
\end{align}
where $\Delta^{\mu\nu}=g^{\mu\nu}-u^\mu u^\nu$. Defining $\nabla_\mu=\Delta_{\mu\nu}\partial^\nu$, one
has:
\begin{align*}
    \theta&=\nabla\cdot u = \partial\cdot u,\\
    \sigma_{\mu\nu}&=\frac{1}{2}\left(\nabla_\mu u_\nu +\nabla_\nu u_\mu\right)-\frac{1}{3}\theta\Delta_{\mu\nu}.
\end{align*} 
By decomposing the tensors $\omega_{\mu\nu}$ and $\Xi_{\mu\nu}$ in the formula \eqref{eq:polarization isothermal} according to the above equations 
it is possible to identify different contributions 
to the spin polarization vector:
\begin{subequations}
\label{eqs:components pol}
\begin{align}
&S^\mu_{A_\omega} =-\epsilon^{\mu\nu\rho\sigma}p_\sigma\frac{\int_{\Sigma}\di\Sigma\cdot p \,n_F(1-n_F)\; 
A_\nu u_\rho }{8mT_H\int_{\Sigma}\di\Sigma\cdot p \, n_F},\\
&S^\mu_{\omega} =\frac{\int_{\Sigma}\di\Sigma\cdot p \,n_F(1-n_F)\; \left[\omega^\mu u\cdot p -u^\nu 
\omega\cdot p\right] }{4mT_H\int_{\Sigma}\di\Sigma\cdot p \, n_F},\label{eq:S_w}\\
&S^\mu_{A_\Xi} = - \epsilon^{\mu\rho\sigma\tau} \hat t_\rho \frac{p_\tau}{\varepsilon}
\frac{\int_{\Sigma} \di \Sigma \cdot p \; n_F (1 -n_F) \left[  u_\sigma A\cdot p + A_\sigma u\cdot p \right]}
{ 8m T_H \int_{\Sigma} \di \Sigma \cdot p \; n_F},\\
&S^\mu_{\sigma} =- \epsilon^{\mu\rho\sigma\tau} \hat{t}_\rho p_\tau \frac{p^\lambda}{\varepsilon} 
\frac{\int_{\Sigma} \di \Sigma \cdot p \; n_F (1 -n_F) \sigma_{\lambda\sigma}}
{ 4m T_H \int_{\Sigma} \di \Sigma \cdot p \; n_F},\\
&S^\mu_{\theta} =- \epsilon^{\mu\rho\sigma\tau} \hat{t}_\rho p_\tau \frac{p^\lambda}{\varepsilon} 
\frac{\int_{\Sigma} \di \Sigma \cdot p \; n_F (1 -n_F)\theta\Delta_{\lambda\sigma}}
{ 12m T_H \int_{\Sigma} \di \Sigma \cdot p \; n_F}.
\end{align}
\end{subequations}
%
\begin{figure}
    \centering
\includegraphics[width=0.49\textwidth]{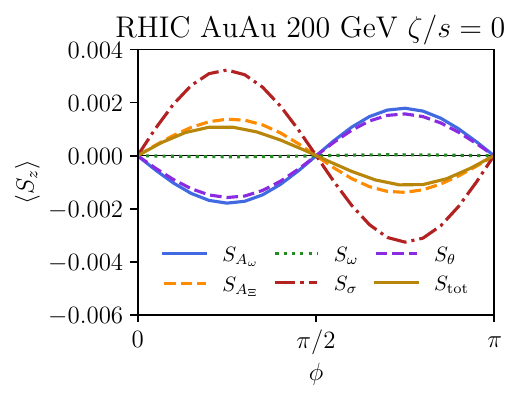}
\includegraphics[width=0.50\textwidth]{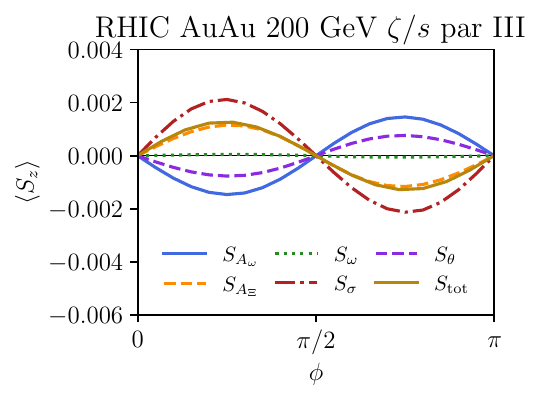}\\
\includegraphics[width=0.49\textwidth]{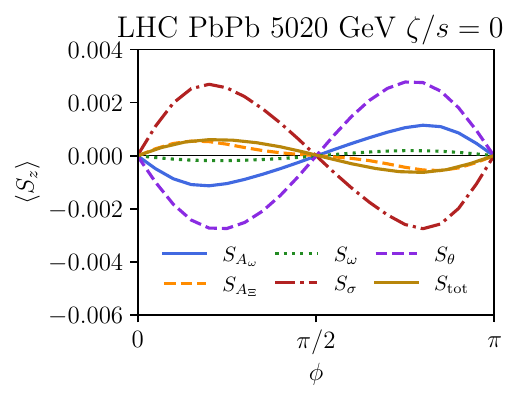}
\includegraphics[width=0.50\textwidth]
{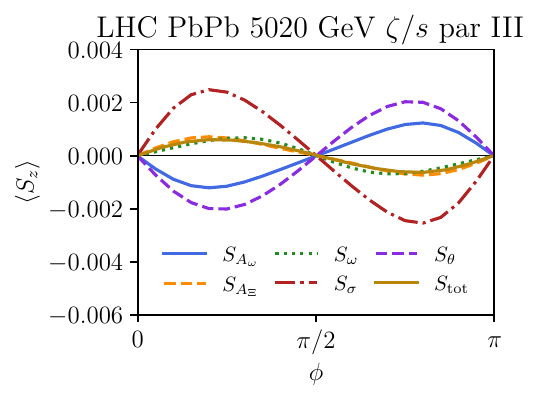}
    \caption{Contributions to the spin polarization vector (see eq.~\ref{eqs:components pol}), 
    for primary $\Lambda$'s at RHIC (centrality window 20-60\%), 
    and LHC energy (centrality window 30-50\%).}
    \label{fig:components pol}
\end{figure}
We have computed the above components for two bulk viscosities: parametrization III of $\zeta/s$ 
in equation \eqref{eq:param3} and for vanishing $\zeta/s=0$, see figure \ref{fig:components pol}. 
In Au-Au collisions at RHIC energy, the contribution from angular velocity is almost vanishing, 
according to previous findings \cite{Karpenko:2018erl} and contrary to the naive lore, which identifies 
rotation as the main source of polarization. Overall, for parametrization III, the amplitude of both $S_\sigma$ and 
$S_\theta$ is smaller than for $\zeta=0$ case, but their sum is approximately zero and 
the total polarization is almost unaffected. On the other hand, in Pb-Pb collisions at LHC energy, 
the situation is more complicated. Without bulk viscosity, $S_\theta$ seemingly cancels $S_\sigma$, 
and $S_{A_\Xi}$ dictates the (negative) sign of the polarization harmonic. Turning on the bulk 
viscosity, not only do $S_\theta$ and $S_\sigma$ get smaller in magnitude, but also $S_\omega$ and 
$S_{A_\Xi}$ significantly change, resulting in the change of sign of the oscillation pattern.

\section{Conclusions}

In summary, we have presented an analysis of spin polarization of $\Lambda$ produced in Au-Au collisions at $\sqrts=200$ GeV and in Pb-Pb collisions at $\sqrts=5020$ GeV, using two different initial state models.
By assuming that the particlization hypersurface is isothermal and including the feed-down corrections,
we have found a good agreement between the data and the predictions of the hydrodynamic-statistical 
model, thus confirming the previous finding \cite{Becattini:2021iol} that isothermal assumption is an
important point to reproduce the data. 
Calculations with \texttt{GLISSANDO} initial state model are in good agreement with the data while 
\texttt{superMC} initial state model reproduces the longitudinal component and the global transverse polarization but fails to reproduce the transverse polarization as a function of the azimuthal angle. 
Transverse polarization is very sensitive to the initial longitudinal flow, and our 
results seem to favor a scenario with initial boost-invariant flow velocity.

We have shown that the longitudinal polarization is very sensitive to bulk viscosity at the highest 
collision energy and almost insensitive to the shear viscosity. At the LHC energy, the presence of 
bulk viscosity flips the sign of the polarization along the beam direction compared to a scenario 
with ideal fluid evolution. 

Altogether, these results demonstrate that spin polarization, and in particular its azimuthal angle 
dependence, can be used as a very effective probe of initial conditions and transport coefficients 
of the Quark-Gluon plasma. 

\section*{Acknowledgements}

We are greatly indebted with C. Shen for his help with the comparison of our results with his. 
This work is supported by ICSC – {\it Centro Nazionale di Ricerca in High Performance Computing, Big 
Data and Quantum Computing}, funded by European Union – NextGenerationEU and by the project PRIN2022
{\it Advanced Probes of the Quark Gluon Plasma} funded by ''Ministero dell'Università e della Ricerca". 
A.P. is supported by the U.S. Department of Energy under Grants DE-FG88ER40388. I.K. acknowledges 
support by the Czech Science Foundation under project No.~22-25026S. Computational resources were provided by the e-INFRA CZ project (ID:90254), supported by the Ministry of Education, Youth and Sports of the Czech Republic.

\bibliographystyle{apsrev4-1}
\bibliography{biblio}

\begin{thebibliography}{67}
\bibitem [{SMASH-vHLLE hybrid with Python()}]{repo}%
  \BibitemOpen
  SMASH-vHLLE hybrid with Python,\ \href@noop {} {}\bibinfo {note} {The
  repository will be made available upon publication.}\BibitemShut {Stop}%
\end{thebibliography}
\end{document}